\documentclass[a4paper,12pt]{iopart}
\usepackage[]{amssymb}
\usepackage[]{amsthm}
\usepackage[]{mathrsfs}
\usepackage[]{tensor}
\usepackage[]{bm}
\usepackage[]{eucal}
\newcommand{\dd}{\partial}
\newcommand{\Lie}{\pounds}
\newcommand{\nab}[1]{\nabla_{\!#1}}
\newcommand{\qqd}{\ , \quad}
\newcommand{\bc}{\begin{center}}
\newcommand{\ec}{\end{center}}
\newcommand{\be}{\begin{equation}}
\newcommand{\ee}{\end{equation}}

\theoremstyle{plain} \newtheorem{tm}{Theorem}[]
\theoremstyle{plain} \newtheorem{lm}[tm]{Lemma}
\theoremstyle{definition} \newtheorem{defn}[tm]{Definition}
\newcommand{\btm}{\begin{tm}}
\newcommand{\etm}{\end{tm}}
\newcommand{\blm}{\begin{lm}}
\newcommand{\elm}{\end{lm}}
\newcommand{\bdefn}{\begin{defn}}
\newcommand{\edefn}{\end{defn}}
\newcommand{\prf}[1]{\noindent \textbf{Proof.} #1 \qed}

\begin{document}

\begin{flushright}
ZTF-EP-14-11
\end{flushright}

\title[Symmetry Inheritance and Jebsen-Birkhoff Theorem]{Symmetry Inheritance and Jebsen-Birkhoff Theorem}

\bigskip

\author{Benjamin Mesi\'c and Ivica Smoli\'c}

\address{Department of Physics, Faculty of Science, University of Zagreb, p.p.~331, HR-10002 Zagreb, Croatia}

\eads{\mailto{benjaminfuture@gmail.com}, \mailto{ismolic@phy.hr}}

\vspace{20pt}

\begin{abstract}
It is known that the Jebsen-Birkhoff theorem is valid for vacuum solutions to Einstein's equation, as well as some of its generalizations. Using symmetry inheritance properties we investigate in detail the additional constraints that fields have to satisfy in order to allow the Jebsen-Birkhoff theorem in the non-vacuum cases of a wide class of gravitational field equations.
\end{abstract}


\bigskip


\vspace{20pt}

\section{Introduction}

A typical model of classical spacetime is a Lorentzian manifold $(M,g_{ab},\psi)$, containing a matter or gauge field(s) $\psi$, which allows at least one Killing vector field $\xi^a$, such that $\Lie_\xi g_{ab} = 0$. There are two natural questions about the symmetries of the fields one might ask in such a context. The first one is whether the field $\psi$ has to share the same symmetries as the metric of the spacetime. When such a concurrence exists, or more concretely, if $\Lie_\xi \psi = 0$ necessarily holds, then we say that the field $\psi$ \emph{inherits} the symmetry. Another question is whether the presence of some particular isometry group $G_n$ implies via field equations the necessity of strictly larger isometry group. This can be answered by a number of results which are usually collected under the umbrella term Birkhoff's theorem \cite{Birk} or, historically more correctly, Jebsen-Birkhoff theorem (JBT) \cite{Jeb,Deser05,JR06}. There are several versions of the precise statement of JBT \cite{Schmidt12}, among which we focus on the one where the spherically symmetric spacetime necessarily allows at least one additional Killing vector field.

\bigskip

The aim of this paper is twofold: a) to revise the symmetry inheritance of some typical fields and, using these results, b) to find what properties fields have to satisfy in order to allow non-vacuum JBT.

\bigskip

We shall consider general form of the gravitational field equation,
\be\label{eq:ET}
E_{ab} = \kappa T_{ab}[\psi]
\ee
where $T_{ab}$ is the energy-momentum tensor and $\kappa$ is some physical constant. The specific form of the tensor $E_{ab}$ will be specified later in the paper. For example, Einstein's equation with cosmological constant $\Lambda$ is the case with $E_{ab} = G_{ab} + \Lambda g_{ab}$, where $G_{ab}$ is the Einstein's tensor. Throughout the text we shall often use the abbreviation $T \equiv g_{ab} T^{ab}$. All the results will be illustrated with three important examples: the ideal fluid,
\be\label{eq:Tidflu}
T_{ab} = (\rho + p) u_a u_b + p g_{ab} \ ,
\ee
the real scalar field $\phi$ with potential $V$ (e.g.~mass term is given by $V_{\mathrm{mass}} = m^2\phi^2/2$),
\be\label{eq:Tscalar}
T_{ab} = \nab{a}\phi \nab{b}\phi - g_{ab} \left( \frac{1}{2}\,g^{cd}\,\nab{c}\phi \nab{d}\phi + V(\phi) \right) \ ,
\ee
and the electromagnetic field $F_{ab}$,
\be\label{eq:Tem}
T_{ab} = \frac{1}{4\pi} \left( F_{ac} \tensor{F}{_b^c} - \frac{1}{4}\,g_{ab}\,F_{cd}F^{cd} \right) \ .
\ee
We assume that the dimension of the spacetime is general $D > 1$, except in the case of electromagnetic field, where $D = 4$.

\bigskip

\section{Symmetry inheritance}

Due to the fact that for every Killing vector field $\xi^a$ we have $\Lie_\xi R_{abcd} = 0$ and that the Lie derivative with respect to a Killing vector field and the covariant derivative commute, the following result follows immediately.

\bigskip 

\blm
Let $\xi^a$ be a Killing vector field and
$$E_{ab} = E_{ab}(g_{cd},R_{cdef},\nab{c}R_{defg},\nab{c}\nab{d}R_{efgh},\dots)$$
a polynomial function. Then $\Lie_\xi E_{ab} = 0$.
\elm

\bigskip

So, in every theory with the gravitational field equation (\ref{eq:ET}) and the tensor $E_{ab}$ specified as in the previous lemma, the energy-momentum tensor shares the symmetries of the spacetime, i.e.~$\Lie_\xi T_{ab} = 0$ is valid for each Killing vector field $\xi^a$. One can now use this piece of information to deduce something about the symmetries of the fields.

\bigskip

In an early analysis by Hoenselaers \cite{Hoen78} it has been proven that the symmetry is inherited by the ideal fluid,
\be
\Lie_\xi \rho = \Lie_\xi p = 0 \quad \textrm{and} \quad \Lie_\xi u^a = 0 \ ,
\ee
as well by the real scalar field with the simplest potential $V_{\mathrm{mass}}$. The letter result can be easily generalized using several tricks. Since the general potential $V$ can be expressed as
\be
V(\phi) = -\frac{T}{D} \pm \frac{D-2}{2D} \sqrt{ \frac{D T_{ab}T^{ab} - T^2}{D-1} } \ ,
\ee
and by assumption $\Lie_\xi T_{ab} = 0$, it follows that
\be
0 = \Lie_\xi V(\phi) = \frac{dV(\phi)}{d\phi}\,\Lie_\xi \phi \ .
\ee
Therefore, at every point where $V'(\phi) \ne 0$ we have $\Lie_\xi \phi = 0$. The case of the massless real scalar field, such that $V(\phi) = 0$, demands a different approach. Here we have
\be
T_{ab} = \nab{a}\phi \nab{b}\phi + \frac{T}{D-2}\,g_{ab} \ ,
\ee
so that
\be
0 = \xi^a \xi^b \Lie_\xi T_{ab} = \Lie_\xi \left( (\Lie_\xi \phi)^2 \right) \ .
\ee
The last equation implies that $\Lie_\xi \phi$ is constant along the orbits (integral curves) of $\xi^a$. In other words, field $\phi$ is a linear function of a parameter along these curves. Hence, assuming that the orbits don't run into a singularity, the exact symmetry inheritance, $\Lie_\xi \phi = 0$, will occur if these curves are compact (topological circles) or if $\phi$ is bounded. An example of unbounded scalar field, a linear function of time in a stationary spacetime, appears in the Case II of Wyman's solution \cite{Wyman81}. We note in passing that a symmetry is not necessarily inherited by the complex scalar field, as was shown by the recent discovery of the scalar hair on Kerr black hole \cite{HR14}. In fact, this ``noninheritance'' lies at the very heart of the circumvention of the Bekenstein's no-hair theorems \cite{Beken72}, which always assume the symmetry inheritance of the fields in the proofs. 

\bigskip

Detailed analysis of the symmetry inheritance for the electromagnetic fields in general relativity has been done by Michalski and Wainwright \cite{MW75}, proving that generally $\Lie_\xi F_{bc} = -a\,{*F}_{bc}$, where for non-null electromagnetic fields $a$ is a constant. Another, more elegant, spinorial proof, independent of particular gravitational field equation, was given more recently by Tod \cite{Tod06}. In the case of $SO(3)$ isometry group, each constant $a_i$, corresponding to Killing vector fields generating this group, is necessarily zero \cite{MW75,Hoen78}. Although there are various conditions which imply inheritance of the stationarity by the electromagnetic field \cite{MW75,Tod06}, in order to avoid any unnecessary additional assumptions, as well as to obtain other information about the nonvanishing components of the field, we show in the Appendix A that all this can be alternatively established by utilizing the spherical symmetry.

\bigskip

\section{Which fields admit Jebsen-Birkhoff theorem?}

Using standard coordinates for a general spherically symmetric $D$-dimensional space\-time,
\be
x^0 = t \qqd x^1 = r \qqd x^i = \theta^i \quad (i=2,3,\dots,D-1)
\ee
where
$$0 \le \theta^i < \pi \quad \textrm{for} \quad i = 2,\dots,D-2 \quad \textrm{and} \quad 0 \le \theta^{D-1} < 2\pi \ ,$$
its metric can be written in the following form (see \cite{HE}, Appendix B)
\be\label{eq:metric}
ds^2 = -e^{2\alpha(t,r)} dt^2 + e^{2\beta(t,r)} dr^2 + \gamma^2(t,r) \, d\Omega_{D-2}^2
\ee
The components of the metric of the spherically symmetric subspace are given by $g_{ii} = \gamma^2\Pi(i)$, where $\Pi(i)$ is the auxiliary function
\be
\Pi(i) \equiv \left\{ \begin{array}{ll} 1 \, , & i = 2 \\ \prod_{k=2}^i \sin^2\theta^k \, , & i \ge 3 \end{array} \right.
\ee
In order to distinguish between different subsets of the general indices $\{\mu,\nu,\dots\}$, we use upper case letters from the beginning of the Latin alphabet $\{A,B,\dots\}$ to denote coordinates from the $t$--$r$ subspace, and lower case letters from the middle of the Latin alphabet $\{i,j,\dots\}$ to denote coordinates from the spherically symmetric subspace. When we speak of a ``diagonal tensor'' $Z_{ab}$ in this particular coordinate system, the phrase can be easily put in a covariant form as $Z_{ab} e^b_{(\mu)} e^c_{(\nu)} = 0$ for each $\mu \ne \nu$, where $\{e^a_{(\mu)}\}$ are the corresponding vielbeins. Finally, we use dot $\,\dot{}\,$ and prime $\,'\,$ to denote derivatives with respect to, respectively, $t$ and $r$ coordinates. It is straightforward to check that the only nonvanishing components of the Riemann tensor for the metric (\ref{eq:metric}) are
\be\label{eq:Riem1}
R_{0101} = \left( \alpha'' + \alpha'^2 - \alpha'\beta' \right) e^{2\alpha} - \left( \ddot{\beta} + \dot{\beta}^2 - \dot{\alpha}\dot{\beta} \right) e^{2\beta}
\ee

\be\label{eq:Riem2}
R_{0i0i} = \left( \alpha'\gamma' e^{2(\alpha-\beta)} + \dot{\alpha}\dot{\gamma} - \ddot{\gamma} \right) \gamma\,\Pi(i)
\ee

\be\label{eq:Riem3}
R_{0i1i} = \left( \alpha'\dot{\gamma} + \dot{\beta}\gamma' - \dot{\gamma}' \right) \gamma\,\Pi(i)
\ee

\be\label{eq:Riem4}
R_{1i1i} = \left( \beta'\gamma' - \gamma'' + \dot{\beta}\dot{\gamma} e^{2(\beta-\alpha)} \right) \gamma\,\Pi(i)
\ee

\be\label{eq:Riem5}
R_{ijij} = \left( 1 - \gamma'^2 e^{-2\beta} + \dot{\gamma}^2 e^{-2\alpha} \right) \gamma^2 \Pi(i) \Pi(j)
\ee

\bigskip

Careful analysis of JBT has to take care about the type of the vector $\nabla^a\gamma$. However, since the different possible cases are in principle analogous \cite{HE}, we restrict our discussion on the case when $\nabla^a\gamma$ is a spacelike vector. This implies that the coordinates can be redefined so that $\gamma = r$. There have been various earlier attempts to characterize energy-momentum tensor which implies JBT \cite{BM95,Faraoni10,FN}. For example, Frolov and Novikov present an elegant sufficient condition.

\bigskip  

\btm
JBT for a solution to Einstein's equation is valid if $T_{AB} = f g_{AB}$ for some function $f$.
\etm

\bigskip

It is interesting to note that this condition is identical to the one which in general relativity implies $g_{00} g_{11} = -1$, as noted by Jacobson \cite{Jacobson07}. More generally, it can be also seen as an ``partial'' equation of state, $T_{00} = -e^{2(\alpha-\beta)} T_{11}$ with vanishing radial momentum density $T_{01} = 0$. 

\bigskip

What are the implications of the condition from the Theorem 2 on the fields? For the ideal fluid, the ``$01$'' component implies that $\rho + p = 0$ or $u_0 = 0$ or $u_1 = 0$. If $\rho + p \ne 0$ then the remaining, $tt$ and $rr$ equations imply that $u_0 = u_1 = 0$, which is in a contradiction with the normalization $u^a u_a = -1$. In conclusion, we necessarily have $\rho + p = 0$ (the cosmological constant case). Symmetry inheritance of the real scalar field allows us to conclude that $\phi = \phi(t,r)$, while the condition from the Theorem 2 puts even stronger further constraint, namely $\phi$ has to be a constant. In the case of electromagnetic field we have the vanishing of the radial component of the Poynting vector (\emph{a posteriori}, JBT and symmetry inheritance imply that the electric and the magnetic fields are spherically symmetric and static). 

\bigskip

In order to investigate whether these constraints are too stringent, it is important to find \emph{necessary and sufficient} conditions for the validity of JBT. Pavelle \cite{Pavelle78} has provided an answer for general relativity in a form of the following theorem.

\bigskip

\btm 
JBT for a spherically symmetric solution to Einstein's equation is valid if and only if the energy-momentum tensor is stationary ($t$-independent) and diagonal.
\etm

\bigskip

We shall argue that Pavelle's theorem can be generalized to a much wider class of theories. From now on, we assume that $E_{ab}$ is a polynomial function formed by the contractions of the Riemann tensors (without covariant derivatives),
\be\label{eq:EpolyR}
E_{ab} = \sum_k E_{ab}^{(k)}(g_{cd}, R_{cdef})
\ee
where $E_{ab}^{(k)}$ denotes a ``monomial'' term. For example, Lovelock has proved \cite{Love71,Love72} that the most general symmetric, divergence free tensor $E_{ab}$, function of metric and its first two derivatives, is exactly of this form.

\bigskip

\blm\label{lm:Ediag}
If $g_{ab}$ is a spherically symmetric static metric and $E_{ab}$ of the form (\ref{eq:EpolyR}), then $E_{\mu\nu}$ is diagonal.
\elm

\medskip

\prf{
One can easily check from the equations (\ref{eq:Riem1}--\ref{eq:Riem5}) that in the stationary case the only nonvanishing components of the Riemann tensor are of the form $R_{\mu\nu\mu\nu}$, up to symmetries. Let us examine an off-diagonal component of the monomial term, $E_{\rho\sigma}^{(k)}$ for $\rho \ne \sigma$. Due to the structure of the components of the Riemann tensor, we know that there would have to be odd number of $\rho$'s among the contracted indices. On the other hand, contracted indices come in pairs, which means that there has to be even number of $\rho$'s. So, the off-diagonal components of the $E_{\mu\nu}$ tensor vanish.
}

\bigskip

It is important to note that this wouldn't be necessarily true in the presence of covariant derivatives of the Riemann tensors. The notable exceptions are the gravitational Chern-Simons terms, which identically vanish for spherically symmetric metrics \cite{BCPDPS11,BCPDPS13}. 

\bigskip

\blm\label{lm:dotbeta}
Let $g_{ab}$ be a spherically symmetric metric. Then $E_{01} \sim \dot{\beta}$.
\elm

\medskip

\prf{
Let assume that a monomial term $E_{01}^{(k)}$ doesn't contain the $R_{0i1i}$ component. Then, due to the structure of the components of the Riemann tensor, we know that there would have to be odd number of ``$0$'' indices among the contracted ones. However, contracted indices come in pairs and hence every monomial term has to contain the $R_{0i1i}$ component. In the $\gamma = r$ case we have $R_{0i1i} \sim \dot{\beta}$.
}

\bigskip

Using these two lemmas, we shall present the scheme by which it is possible to generalize Pavelle's theorem in the following form,

\begin{quote}
\textbf{Proposition.} Suppose that JBT is valid in the vacuum case of (\ref{eq:ET}). Then JBT in the presence of fields is valid if and only if the energy-momentum tensor is stationary and diagonal.
\end{quote}

\noindent
The reason why this statement is not present as an rigorous theorem is related to the generality of the claim and will be commented at the end of the discussion that follows.

\bigskip

Suppose that JBT is valid for (\ref{eq:ET}). Then the metric $g_{ab}$ is stationary, which implies that $E_{ab}$ is stationary and hence the energy-momentum tensor is stationary too. Also, from the Lemma \ref{lm:Ediag} we know that $E_{\mu\nu}$ is diagonal, so is $T_{\mu\nu}$. Conversely, let us suppose that the energy-momentum tensor is stationary and diagonal. From the Lemma \ref{lm:dotbeta} we know that $E_{01} \sim \dot{\beta}$. In the vacuum case this component of the field equation is used to conclude that $\dot{\beta} = 0$ and hence $g_{11,0} = 0$. Since the energy-momentum tensor is diagonal, the off-diagonal components of the field equations remain unchanged in the non-vacuum case, so that this conclusion remains unaltered. Furthermore, using $\beta = \beta(r)$ and $\gamma = \gamma(r)$ we see that in the components of the Riemann tensor remain only derivations with respect to the $r$ coordinate. Since the components of the energy-momentum tensor are $t$-independent, the diagonal components of the field equations allow solution for the $g_{00}$ component of the form $g_{00} = p(t)q(r)$. Then the $p$ function can be used to redefine coordinate $t$ and hence $g_{00,0} = 0$. In conclusion JBT remains valid in the non-vacuum case under given assumptions.

\bigskip

One could possibly contrive an example of gravitation field equation where the nonvacuum JBT doesn't hold because the conclusion $\dot{\beta} = 0$ depends on ``diagonal equations'' in such a way that the presence of $r$-dependent terms $T_{\mu\mu}$ ruins its validity. This, however, doesn't happen in various important extensions of general relativity where JBT is known to be valid in the vacuum case, such as the Lovelock's gravity \cite{Whitt88,Zegers05} and Palatini $f(R)$ theory \cite{Faraoni10}. Thus, a possible ``pathological'' case of tensor (\ref{eq:EpolyR}) will violate usual fundamental physical prerequisites (assumptions from the Lovelock's theorem) for the gravitational field equation. It remains to be seen to what extent this proposition can be extended if the covariant derivatives of Riemann tensors are present in the tensor $E_{ab}$.

\bigskip

Finally, let us investigate consequences of the conditions from generalized Pavelle's theorem and symmetry inheritance properties. For the ideal fluid we get $\rho = \rho(r)$, $p = p(r)$ and either $\rho + p = 0$ or $u^0 = u^0(r)$ as the only nonvanishing component of $u^a$. Similarly, for the real scalar field we get either $\phi = \phi(t)$ or $\phi = \phi(r)$ (\emph{a posteriori}, the former option is incompatible with JBT and symmetry inheritance). In the presence of the black hole horizon, due to Bekenstein's no-hair result \cite{Beken72}, the field $\phi$ must be constant. Diagonal electromagnetic energy-momentum tensor implies the vanishing of the Poynting vector (see e.g.~\cite{Heusler}, section 5.1), so that $E^a$ and $B^a$ are parallel, which is consistent with the conclusions from the symmetry inheritance.

\bigskip

\ack

This work was partially supported by the Croatian Ministry of Science, Education and Sport under the contract No.~119-0982930-1016.

\bigskip

\appendix

\section{Stationarity inheritance of the electromagnetic field}

Suppose that the spacetime is spherically symmetric and that due to JBT it is stationary too, with corresponding Killing vector field $k^a = (\dd/\dd t)^a$. Then one can introduce \cite{Heusler,Smolic14} electric and magnetic 1-forms, $E = -i_k F$ and $B = i_k {*F}$. Let $\xi^a$ be a Killing vector field such that $\Lie_\xi F_{ab} = 0$ and $\Lie_k \xi^a = 0$. Then, using the identity
\be
\Lie_X i_Y - i_Y \Lie_X = i_{[X,Y]}
\ee
and the fact that the Lie derivative commutes with Hodge operator, we have
\be
\Lie_\xi E = -\Lie_\xi i_k F = -i_k \Lie_\xi F = 0
\ee
\be
\Lie_\xi B = \Lie_\xi i_k {*F} = i_k \Lie_\xi {*F} = 0
\ee
Since $\xi^a$ is a Killing vector field, the same is true for dual vector fields $E^a$ and $B^a$, $\Lie_\xi E^a = \Lie_\xi B^a = 0$. More concretely, using the fact that the non-null electromagnetic field inherits the spherical symmetry and that well known generators of the $SO(3)$ isometry (see e.g.~\cite{Carroll}, equation (5.25)) commute with $k^a$, these equations imply that the only nonvanishing components of the electric and magnetic fields are $E^1 = E^1(t,r)$ and $B^1 = B^1(t,r)$. Correspondingly, the only nonvanishing components of $F_{ab}$ are thus $F_{01} = F_{01}(t,r)$ and $F_{23} = f(t,r)\sin\theta$ for some function $f$. Covariant form of the vacuum Maxwell's equations can be written as
\be
\nabla^\mu F_{\mu\nu} = \frac{1}{\sqrt{-g}}\,g^{\mu\rho} \dd_\rho \left( \sqrt{-g} \, F_{\mu\nu} \right) = 0
\ee
\be
\nabla^\mu\,{*F}_{\mu\nu} = \frac{1}{\sqrt{-g}}\,g^{\mu\rho} \dd_\rho \left( \sqrt{-g} \, {*F}_{\mu\nu} \right) = 0
\ee
where $g$ is the determinant of the metric. Using the information we have obtained about the components of $F_{ab}$, $\nu = 1$ components of the Maxwell's equations allow us to finally deduce that $F_{01}$ and $F_{23}$ are in fact $t$-independent in a spherically symmetric static spacetime. In conclusion, under given assumptions, the electromagnetic field inherits the stationarity, $\Lie_k F_{ab} = 0$.

\bigskip

In such a context, one can also introduce \cite{Heusler,Smolic14} locally defined electric scalar potential $\Phi$ and the magnetic scalar potential $\Psi$, defined via $E = d\Phi$ and $B = d\Psi$. Since the Lie derivative and the exterior derivative commute, it follows that $\Lie_\xi \Phi$ and $\Lie_\xi \Psi$ are both constant. Furthermore, in the presence of Dirichlet boundary conditions on the hypersurfaces invariant under the action of a Killing vector field $\xi^a$ (e.g.~the electromagnetic potentials are constant on Killing horizons \cite{Smolic12,Smolic14}) one can deduce that both $\Lie_\xi \Phi$ and $\Lie_\xi \Psi$ are in fact zero. In the case of spherically symmetric static spacetime this implies that $\Phi = \Phi(r)$, $\Psi = \Psi(r)$, and thus the electric and magnetic fields are spherically symmetric, static and radial, $E = E(r)\,dr$ and $B = B(r)\,dr$, in agreement with the conclusions from above.

\section*{References}

\bibliographystyle{unsrt}
\bibliography{birkmattref}

\end{document}